\documentclass[12pt,a4paper]{article}

\usepackage{graphicx,here}
\usepackage{bm}

\begin{document}

\begin{center}

{\Large Phase transition of the dissipative double-well quantum mechanics}

\vspace{5mm}

{\small Ken-Ichi Aoki \footnote{aoki@hep.s.kanazawa-u.ac.jp} and Tamao Kobayashi \footnote{kobayasi@yonago-k.ac.jp}}

\vspace{5mm}

{\small \it $1$)Institute for Theoretical Physics, Faculty of Mathematics and Physics, 
Kanazawa University, \\Kanazawa, Ishikawa, Japan\\
and\\
$2$)General Education, 
Yonago National College of Technology, \\Yonago, Tottori, Japan}

\vspace{5mm}

{\bf \small abstract}
\end{center}

{\small We investigate the critical dissipation of the double-well quantum mechanics.
We adopt two-state approximation to define effective Ising models and apply
the block decimation renormalization group and 
the finite range scaling method recently proposed for the long range 
Ising model. We briefly report the numerical results of the critical dissipation
for various model parameters.}

\vspace{10mm}

\section{Introduction}
The phase transition here is the quantum-to-classical phase transition.
Consider a quantum system of the double well potential. Due to the tunneling,
the system is oscillating and symmetric, which we call `quantum'.
With dissipative effects, the tunneling is suppressed, 
and the ground state becomes a localized state to break the original Parity symmetry, 
which we call `classical'.
Then, there must be a `phase transition' between these two states.
Our subject is to evaluate the critical dissipation for this phase transition using
a new renormalization group method and a new scaling ansatz according to the
range of interactions.

The above picture about the phase transition
has been common now by many researches with various methods
\cite{Caldeira-Leggett,Instanton-DW,MC-DW,suzuki92}. 
This type of transition is expected to be observed as decoherence
effects in the recent quantum mechanical few body experiments\cite{Exp}
or future quantum devices. Quantitative treatment of decoherence processes
is essential in planing realistic quantum devices, but it 
must be a difficult subject since it need to bridge between the quantum and
classical notions. Certainly the double-well quantum mechanics 
with dissipative effects is a first step model which we must fully understand
quantitatively. 

The actual value of the criticality for the model as a function of 
the parameters of the double-well potential and the dissipative interactions, 
are not well established.
Also, the double-well system has been frequently compared to the one-dimensional
Ising model especially for the deep well case. However, the detailed and quantitative
relations between the double-well quantum mechanics and the Ising model with
general type of interactions have not been clear yet.

In this article, we develop new methods of handling the infinite range
interactions, that is, the block decimation renormalization group and the finite range
scaling. These two methods are new way of evaluating the infinite range
system with recourse to the intermediate finite range systems.
The block decimation renormalization group is a real space renormalization group
method to solve finite range system exactly.
Then the finite range scaling ansatz is applied to the susceptibility 
calculated exactly for finite range systems, and we evaluate its scaling 
exponent. The finite range scaling exponent finally tells us the value of the
critical coupling constant through the pole position of the zeta function.
These new methods may shed light on other infinite range systems 
even in higher dimensional space, and may open new analyzing power in the 
renormalization group world.

First of all we should recall that 
it is non-trivial to include dissipative effects in the quantum mechanics, since
there is no simple Hamiltonian to realize friction proportional to the velocity.
We take the Caldeira-Leggett (CL) model\cite{Caldeira-Leggett}, 
where dissipation comes out of the microscopic origin.
The model consists of a target system
and environmental degrees of freedom of infinitely many harmonic oscillators,
\begin{equation}
{S}[~q,\{x_{\alpha}\}]
=
\int d{t}~\left\{\ 
\frac{1}{2}M{\dot q}^{2} - V_0(q)
\right. + \sum_{\alpha}
\left.\left[~\frac{1}{2}m_{\alpha}{\dot x_{\alpha}}^{2}
-\frac{1}{2}m_{\alpha}\omega_{\alpha}^{2}x_{\alpha}^{2}
-q C_{\alpha}x_{\alpha}\right]\ \right\}\ .
\end{equation}
The target variable $q$ and each environmental variable $x_\alpha$ are coupled linearly 
with coupling constant $C_\alpha$.
Due to these couplings, the energy of the target system is transfered
to the environmental oscillators, and it exhibits as dissipative effects.

To describe the effective target dynamics, we path integrate out the environmental 
degrees of freedom. 
Then the non-local interactions of the target variable emerge as,
\begin{eqnarray}
\Delta S_{\rm NL}=\frac{\eta}{4\pi} \int {\rm d}s {\rm d}\tau 
\ {(q(s) -q(\tau))^2 \over |s-\tau|^p}\ ,
\end{eqnarray}
where $\eta$ is an effective coupling constant to represent the strength of dissipation
and parameter $p$ denotes the damping rate of the non-local interactions.
Of course these parametrization is a simplified expression of non-local
interactions which are 
determined by micro parameters $m_\alpha, \omega_\alpha$ and $C_\alpha$.
It is known that with quadratic damping ($p=2$) interactions, 
the classical equation of motion of $q(t)$ suffers effectively a friction proportional to 
the velocity $\displaystyle {dq \over dt}$, and thus this case is called the Ohmic case.

In the Euclidean path integral formalism, the quantum mechanical system is equivalent to 
a one-dimensional statistical system. If we approximate the system using the minimum 
degrees of freedom, that is, two states par site (after discretization in time 
with a finite slice), 
it is transformed into the
Ising spin chain with long range interactions.

Long range Ising models with the above type of simple interactions,
\begin{eqnarray}
-\beta H=\sum_k \sum_n K_n \sigma_k \sigma_{k+n}=
\sum_k \sum_n {\eta \over n^p}\sigma_k \sigma_{k+n}\ ,\ \sigma_k =\pm 1\ ,\label{eq:Ising}
\end{eqnarray}
have its own long history pioneered by Refs.\cite{Griffiths67,Ruelle68,Dyson69}, and
it has been understood that with strong enough long range interactions there occurs the 
spontaneous magnetization.
The known facts\cite{Froehlich-Spencer82,Aizenman-Fernandez88,Aizenman--Newman88,Imbrie-Newman} 
about the phase transition are that
the critical coupling constant $\eta_{\rm c}$ is finite only for
$1 < p \le 2$, that is, $\eta_{\rm c} =0$ for $p\le 1$ and $\eta_{\rm c} = \infty $ for $p > 2$.
Particularly the $p=2$ case, which is called as "Ohmic" because it corresponds to the
velocity proportional friction in the classical mechanical situation, 
is a very special boundary where it is proved that the Kosterlitz-Thouless transition occurs
\cite{KT}.

It is difficult, however, to evaluate the value itself of $\eta_{\rm c}$.
In the previous article we successfully evaluated the critical coupling constant 
of this long range Ising model, where we developed a new method of investigating
the infinite range system, Finite Range Scaling (FRS) hypothesis\cite{Aoki-Kobayashi-Tomita08}. 

In this article we apply this method to the double-well quantum mechanics.
Here we briefly report the prototype calculation using two-state approximation of the
system. Even with two-state approximation, the effective Ising spin interactions
are not a straightforward mapping of the original non-local interactions.
We will confirm that FRS method does work to evaluate the critical dissipation and our
results are consistent with those obtained by other methods.
The full calculation without state reduction will be reported elsewhere.

\section{Finite Range Scaling method in the long range Ising model}

\begin{figure}
\begin{center}
\includegraphics[width=9cm]{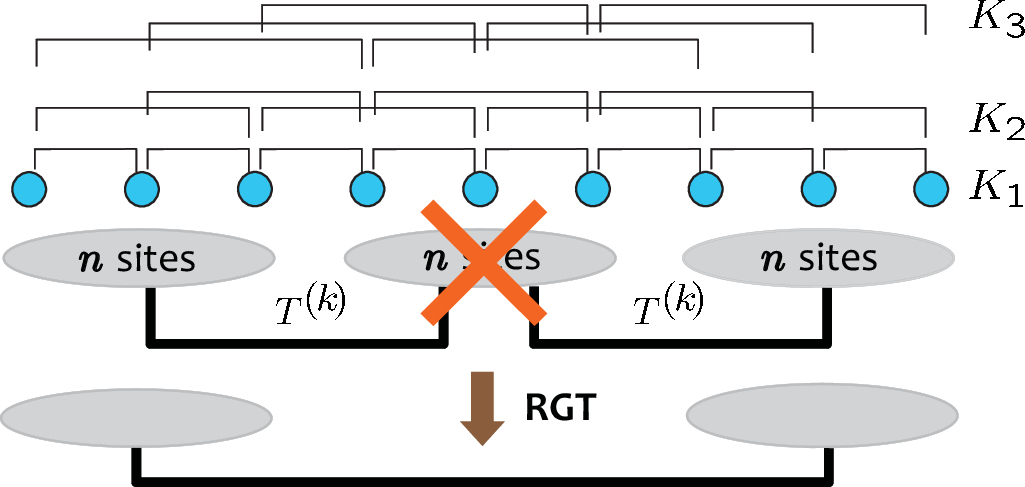}
\caption{BDRG procedure}
\end{center}
\vskip-3mm
\end{figure}

For the long range Ising models defined in Eq.(\ref{eq:Ising}), we successfully evaluated the critical dissipative
interactions to bring about the spontaneous magnetization. 
We developed a new method of the 
Block Decimation Renormalization Group (BDRG) and the Finite Range Scaling (FRS) ansatz
\cite{Aoki-Kobayashi-Tomita08}.
Our method utilizes simple calculation with very small size of the computational
resources.
There are some large scale Monte Carlo studies\cite{MC-old,MC-Ising97,MC-Ising01} 
for the limited parameter space, and 
their results are perfectly consistent with ours, which indicates our ansatz of 
FRS is just on the right way of accessing properties of infinite range interactions. 

We explain the BDRG method in Fig.1 in case of range $n=3$.
The DRG (Decimation Renormalization Group) is a block spin transformation method
formulated by Wilson to calculate the partition function approximately\cite{Wilson75}.
In one dimensional nearest-neighbor Ising system, it works without any approximation and we can get
the exact partition function very easily.
In the case that there are non-nearest-neighbor interactions, 
the original DRG does not work.
We developed an extended version, the Block DRG, which may accommodate long range interactions.

We consider a system with long range interactions with coupling constants $K_m$ which
couples $m$-separated two spins.
We set the maximal finite range of interactions to be $n$, and investigate 
the infinite range limit by increasing $n$, one by one.
First, we divide spins into blocks of size $n$. 
Then there are interactions only between nearest neighbor blocks, that is, 
looking at the model `block-wisely', it is nothing but the nearest-neighbor model where
one block plays a site of $2^n$ states.

Next, all the spins in the middle block in Fig.1 are decimated, namely, are
integrated for all possible $2^n$ states. After decimation we obtain new 
inter-block interactions which still keeps nearest-neighbor-block property.
Now the renormalization transformation is defined and can be repeated without
approximation. 

We represent inter-block interactions by the transfer matrix $T$ of 
$2^n \times 2^n$.
Then the renormalization transformation is nothing but to make
a product of $T$ matrices, 
\begin{equation}
T^{(k+1)} = T^{(k)}\cdot T^{(k)}\ ,
\end{equation}
where $T^{(k)}$ matrix is inter-block interactions between 
$k$-th renormalized neighboring blocks.

By BDRG we calculate the susceptibility of the system $\chi (n)$ with a finite range 
interactions exactly. We impose the extend field $h$ as,
\begin{equation}
-\beta H =\sum_k \left[ \sum_n K_n \sigma_k \sigma_{k+n} +h\sigma_k \right]\ .
\end{equation}
Then the susceptibility of the system is given by,
\begin{equation}
\chi (n) =\lim_{k \to \infty}\frac{1}{n\cdot 2^k} \left. {\partial^2 \left(\ln {\rm Tr} T^{(k)}\right) \over \partial h^2}\right|_{h=0}\ .
\end{equation}

We increase the range $n$ by 1 to see the change of the susceptibility and 
define the corresponding scaling exponent $\beta$ as follows,  
\begin{equation}
\log\chi (n) -\log\chi (n-1) \propto \left( \frac{1}{n} \right)^{\beta(p,\eta)}.
\end{equation}
We expect that this exponent $\beta$ will be asymptotically a constant for 
enough large $n$. This is a hypothesis and we call it Finite Range Scaling (FRS).

If the FRS does work, 
then the infinite $n$ behavior of the susceptibility apart from the finite part
can be evaluated as 
\begin{equation}
\log \chi(n\rightarrow\infty) 
= \sum_n^{\infty}\left[\log\chi (n) -\log\chi (n-1)\right]
\simeq \sum_n^{\infty}\left( \frac{1}{n} \right)^{\beta(p,\eta)}
=\zeta (\beta(p, \eta))\ .
\end{equation}
Using the exponent we estimate the 
divergent point of the susceptibility and it gives the critical dissipation.
That is, the infinite $n$ behavior of the susceptibility is controlled by 
the zeta function $\zeta(\beta)$, which has a pole singularity at $\beta=1$.
Therefore, the critical $\eta$ is determined by the condition $\beta(p,\eta_{\rm c})= 1$.
This method of using finite range systems to evaluate the infinite range divergence is 
the Finite Range Scaling method we have developed.

\section{Effective Ising model of the double-well quantum mechanics}

We move on to the double well quantum mechanics, 
\begin{equation}
V(x) =-\frac{1}{2} x^2 +\lambda x^4.
\end{equation}
Note that the number of states on a site is infinite in quantum mechanics and 
$T$ matrix is replaced by a bi-local function defined by a bi-local potential $W$ as, 
\begin{eqnarray}
T(x,y)=e^{-W(x, y)} =\langle x \mid {\hat U} \mid y\rangle\ .
\end{eqnarray}
The operator $\hat U$ is the imaginary time evolution operator and this
bi-local function is nothing but the Feynman path integral kernel.

Adding the non-local interactions generated by the path integration of environmental degrees of freedom,
we have to compose $n$ sized blocks to make the system `nearest-neighbor'.
Then the inter-block $T$ matrix is a multi variable function with $2n$ variables as follows, 
\begin{eqnarray}
T=e^{-W(x_1 \cdots x_n,\ y_1 \cdots y_n)} =\langle x_1 x_2 \cdots x_n \mid {\hat U} \mid y_1 y_2 \cdots y_n \rangle\ .
\end{eqnarray}
Here we introduce a new complete set of states $\mid a_n\rangle$  
and express $T$ matrix with this new base system,
\begin{eqnarray}
T&=& e^{-{\widetilde W}(a_1 \cdots a_n , b_1 \cdots b_n)} = \langle a_a \cdots a_n \mid {\hat U} \mid b_1 \cdots b_n \rangle \nonumber \\
&=& \int dx_1 \cdots dx_n dy_1 \cdots dy_n \langle a_1 \cdots a_n \mid x_1 \cdots x_n \rangle  \langle x_1 \cdots x_n \mid {\hat U} \mid y_1 \cdots y_n \rangle \nonumber \\
&&\times\langle y_1 \cdots y_n \mid b_1 \cdots b_n \rangle \nonumber \\
&=& \int d x_1 \cdots dx_n dy_1 \cdots dy_n \psi_{a_1}^{\ast} (x_1) \cdots \psi_{a_n}^{\ast} (x_n) e^{-W (x_1 \cdots x_n , y_1 \cdots y_n)} \nonumber \\
&&\times\psi_{b_1} (y_1) \cdots \psi_{b_n} (y_n), 
\end{eqnarray}
where $\psi_a(x)$ is the wave function of state $|a\rangle$.
Our site variables $x,\ y$ are converted into other complete set suffix $a, b$.

Now we approximate the states on a site with only two states, that is, 
the summation in $a,b$'s are just two-fold. This defines an effective
Ising model of one-dimension which has general interactions between spins, 
and we can write the multi-local potential as
 $\widetilde W({\bm \sigma}_a, {\bm \sigma}_b)$ where ${\bm \sigma}_a, {\bm \sigma}_b=\{-1, +1\}$.

The simplest version of two-state approximation 
is to take the completely localized states at the left and right
well bottoms, where the wave functions are the $\delta$-function.
Then the converted $\widetilde W({\bm \sigma}_a, {\bm \sigma}_b)$ 
function is directly related to
the original $W({\bm x},{\bm y})$ function as follows,
\begin{equation}
\widetilde W({\bm \sigma_a, \bm\sigma_b})= W(v{\bm \sigma}_a, v{\bm \sigma}_b)\ ,
\ \   v=\frac{1}{2\sqrt{\lambda}}\ ,
\end{equation}
where $\pm v$ is the position of the double well bottoms.
The effective spin interactions in this case are given by,
\begin{equation}
\epsilon^{2-p}
{\eta \over 4\pi\lambda} \sum_k \sum_n \frac{1}{n^p}\sigma_k \sigma_{k+n}
+ \frac{m}{4\lambda\epsilon}\sum_k \sigma_k \sigma_{k+1}
\ ,
\end{equation}
which is exactly the same form of the normal long range Ising model
defined in Eq.(\ref{eq:Ising}) 
except for the total rescaling of the coupling constant $\eta$ and 
the nearest-neighbor interactions due to the kinetic term of the quantum system.
Then we can easily estimate the critical dissipation by referring to the Ising model
result calculated in Ref.\cite{Aoki-Kobayashi-Tomita08,MC-Ising01} as follows: 
\begin{equation}
\eta_{\rm c} = 4\pi \lambda \times \eta_{\rm c}\mbox{\rm [Ising]}\ ,
\end{equation}
where we assume that
the criticality does not strongly depend on the size of the nearest neighbor term.
This relation holds for any $p$ and in the Ohmic case ($p=2$), for example,  we have
\begin{equation}
\eta_{\rm c} \simeq 4\pi \lambda \times  0.66\ .
\label{eq:delta-wf}
\end{equation}
This result is quite similar in the $\lambda$ dependence 
to that obtained by the dilute instanton 
calculation\cite{Instanton-DW},
\begin{equation}
\eta_{\rm c} = 2\pi \lambda\ ,
\end{equation}
and we note that it is larger than this instanton approximation.

However this selection of states makes difficulty that there appears divergences due to 
the vanishing time discretization slice $\epsilon$. For example the correlation length $\xi$
of the system diverges as
\begin{equation}
\xi \simeq
\frac{\epsilon}{2} \exp\left({\frac{m}{2\lambda} \frac{1}{\epsilon}}\right) .
\end{equation}
because of the diverging nearest neighbor interactions.
This is not consistent with the fact that the quantum mechanics does not need
any renormalization of bare parameters.

Instead, we take linear combination of two states, the ground state $|0 \rangle$ 
and the 1st excited state $|1 \rangle$ of the double well potential 
without dissipation ($\eta=0$),  
$\displaystyle \psi_{\uparrow ,\downarrow} = (|0\rangle \pm |1\rangle) / \sqrt{2}$,
which are regarded as left or right states corresponding to up or down of Ising spin.
We call this type of  two-state approximation as the ground state approximation.
In this approximation the correlation length of the system is evaluated as
\begin{equation}
\xi =-\epsilon \left[ \log \tanh \left( -\frac{1}{2} \log \tanh 
\left( \frac{\epsilon \delta}{2}\right) \right) \right]^{-1},
\end{equation}
where $\delta$ is the energy gap between the ground and the 1st excited states.
At the vanishing time slice, it converges as 
\begin{equation}
\lim_{\epsilon\rightarrow 0}\xi(\epsilon) = \frac{1}{\delta}\ ,
\end{equation}
which is the correct value of the continuum system.

Hereafter, our aim is to evaluate the plausibility of our BDRG and FRS methods
in the effective Ising model with the ground state approximation.
Note that the effective Ising models in this approximation have all possible spin interactions including multi-spin products,
which might be far away from the normal long range Ising interactions in Eq.(\ref{eq:Ising}).
Therefore it is non-trivial that
FRS method works with such general type of long range Ising models.

For example, in $n=2$, initial $T$ matrix is calculated  as follows, 
\begin{eqnarray}
&&T= e^{-{W}(a_1  a_2 ,\  b_1  b_2)} = \int dx_1 dx_2 dy_1 dy_2\ \psi_{a_1}^{\ast}(x_1) \psi_{a_2}^{\ast}(x_2) \psi_{b_1} (y_1) \psi_{b_2} (y_2) \nonumber \\
&& \times \exp \left[ -\frac{m(x_1 - x_2)^2}{4\epsilon}-\frac{m(x_2 -y_1)^2}{2\epsilon} -\frac{m(y_1 -y_2 )^2}{4\epsilon} \right. \nonumber \\
&&\left. -\frac{\epsilon}{2} \left( V (x_1)+ V (x_2) +V (y_1) +V (y_2) \right) \right] \nonumber \\
 &&  -\frac{\eta}{2\pi} \epsilon^{2-p} \left[ \frac{1}{2} (x_1 -x_2)^2 +(x_2-y_1)^2 +\frac{1}{2} (y_1 -y_2)^2 \right. \nonumber \\
&&+\left. \frac{1}{2^p} \left( (x_1 -y_1)^2 + (x_2 -y_2)^2 \right) \right]\ ,
\end{eqnarray}
where $\epsilon$ is the discretization step for the imaginary time and $a\ ,b$ are Ising variables $\{ \uparrow\ , \downarrow\}$.
The first group of terms are the original kinetic terms, the second are the potential terms, 
and the last are dissipation terms given by the CL non-local interactions.

In the ground state approximation, generally,
$2^{2n}$ integrations of $2n$-dimensions are necessary to get the initial 
$T$ matrix of BDRG. To evaluate these large dimensional integration, 
we adopt the Monte Carlo integration method.
We set up random numbers obeying the probability distribution function 
defined by all local terms in the integral, namely, 
wave functions and potential terms. Wave functions are not positive 
semi-definite and we introduce sign functions to take account of the
negative region of the wave functions. Kinetic terms and non-local 
interaction terms are evaluated by random numbers defined above.

After obtaining the effective spin interactions, we modify it so that
the $T$ matrix respects the Parity symmetry and mirror symmetry.
For example, in $n=2$, these symmetries are,
\begin{equation}
W({\bar a}_1 {\bar a}_2\ ,{\bar b}_1 {\bar b}_2)=W(a_1 a_2,\ b_1 b_2),\ W(b_2 b_1\ ,a_2 a_1)=W(a_1 a_2,\ b_1 b_2)\ ,
\end{equation}
where $\displaystyle \overline{\uparrow}=\downarrow\ , \overline{\downarrow}=\uparrow$ .
This symmetrization is important and necessary, since if not, the renormalization 
transformation enhances the symmetry breaking components and the 
resultant renormalized $T$ matrix will be out of reality.
We checked the validity and precision of our Monte Carlo integration by comparing
our results with those obtained by Simpson's integration method which can be done
for very low $n$ cases.

\section{Results}

\begin{figure}[ht]
\begin{center}
\includegraphics[width=0.6\textwidth]{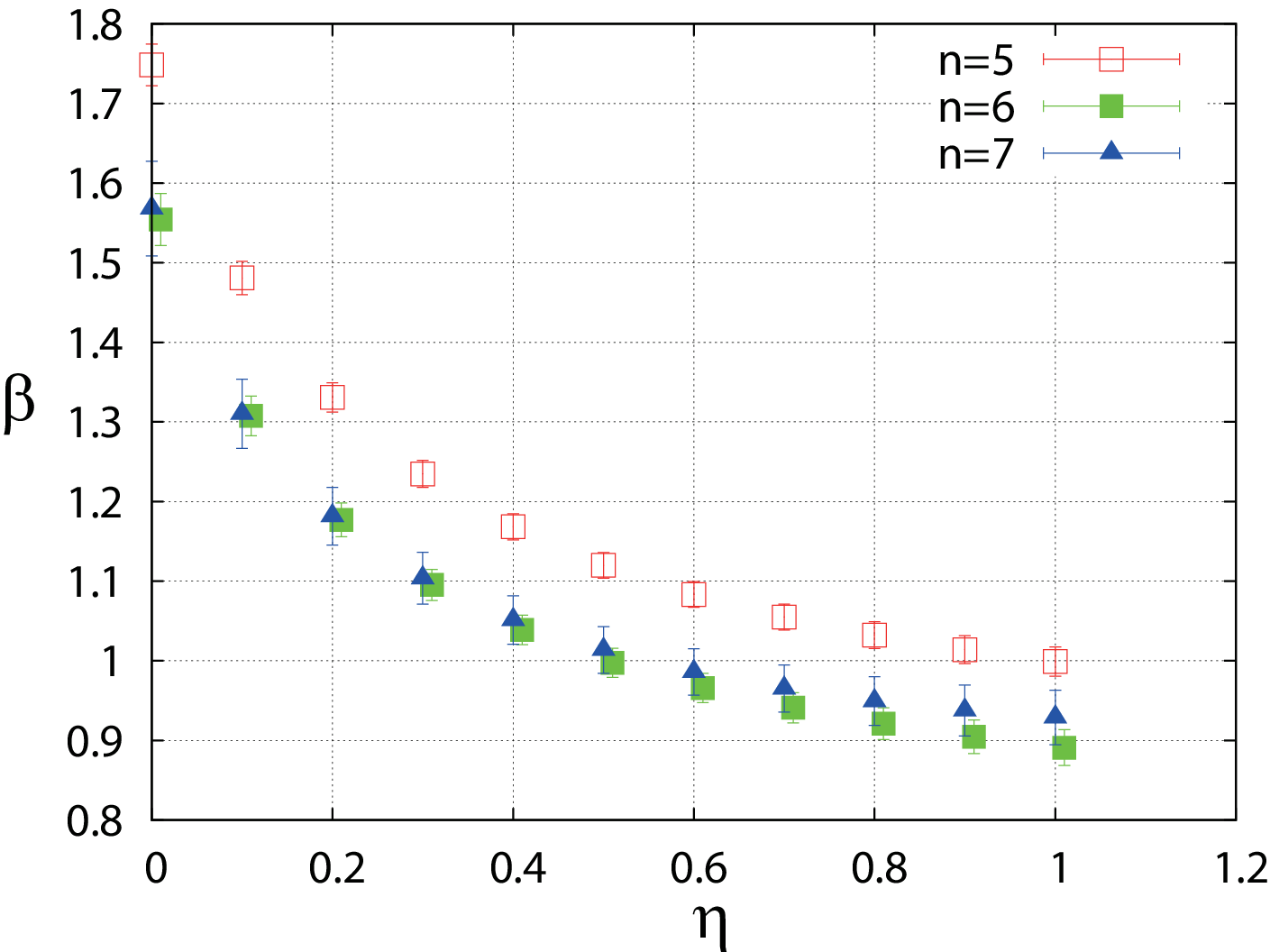}
\caption{Typical behavior of $\beta$.}
\label{fig:beta-behavior}
\end{center}
\vskip-2mm
\end{figure}

We show our numerical results in order.
Fig.\ref{fig:beta-behavior} shows a typical behavior 
of the Finite Range Scaling exponent
$\beta$ versus $\eta$ for $n=5,6,7$. Here we take
 $\lambda=0.04$, the damping rate $p=1.99$, 
the discretization step $\epsilon=0.9$, the number of configurations for Monte Carlo 
integration =$1.28$ million.
We calculated $\beta$ with 16 different set of random numbers and evaluated the
standard deviation of the data which are plotted in the figure as $1\sigma$ bar.
This initial evaluation is called {\sl rough} estimate.

The global behavior of $\beta$ is very smooth and natural.
It decreases monotonically with respect to $\eta$ and crosses the critical
value $\beta=1$. According to the FRS hypothesis, 
the crossing point is the critical coupling constant $\eta_{\rm c}$.
Note that we need estimates of $\beta$ with enough large $n$,  and 
$n=6$ or $n=7$ are actually `large' since it seems they are already stable
against $n$. This exhibits a surprising feature that with relatively short
range system we can evaluate the criticality which can emerge only for
infinite range interactions. Such features of FRS method is common to the 
normal long range Ising models. 

Thus we can obtain the critical coupling constant by FRS method. 
This holds for all cases we calculated with various system parameters 
$\lambda, p, \epsilon$.

\begin{figure}[ht]
\begin{minipage}{0.45\hsize}
\begin{center}
\includegraphics[width=0.99\textwidth]{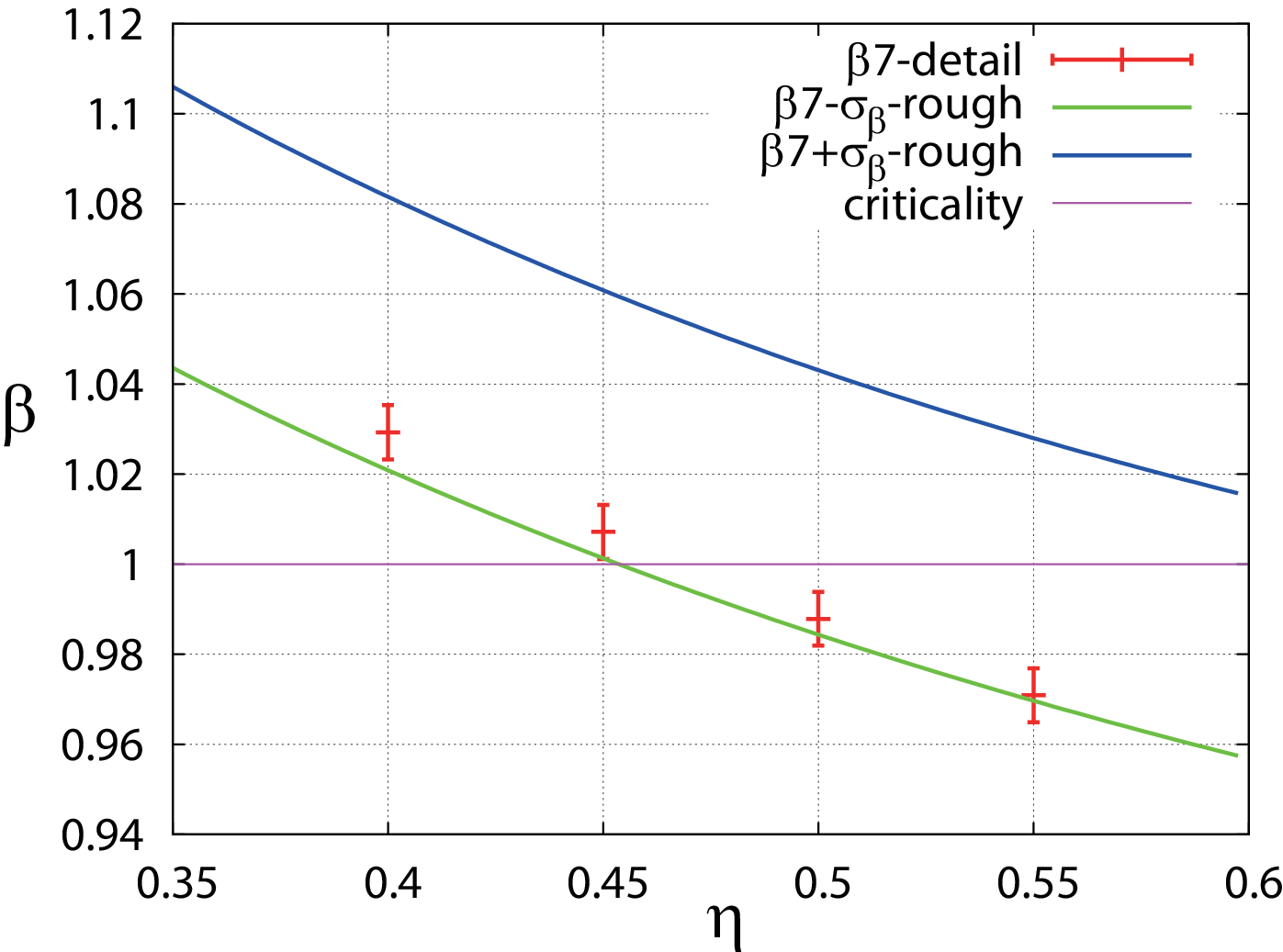}
\caption{Detailed behavior of $\beta$.}
\label{fig:beta-detail}
\end{center}
\end{minipage}
\hfill
\begin{minipage}{0.45\hsize}
\begin{center}
\includegraphics[width=0.99\textwidth]{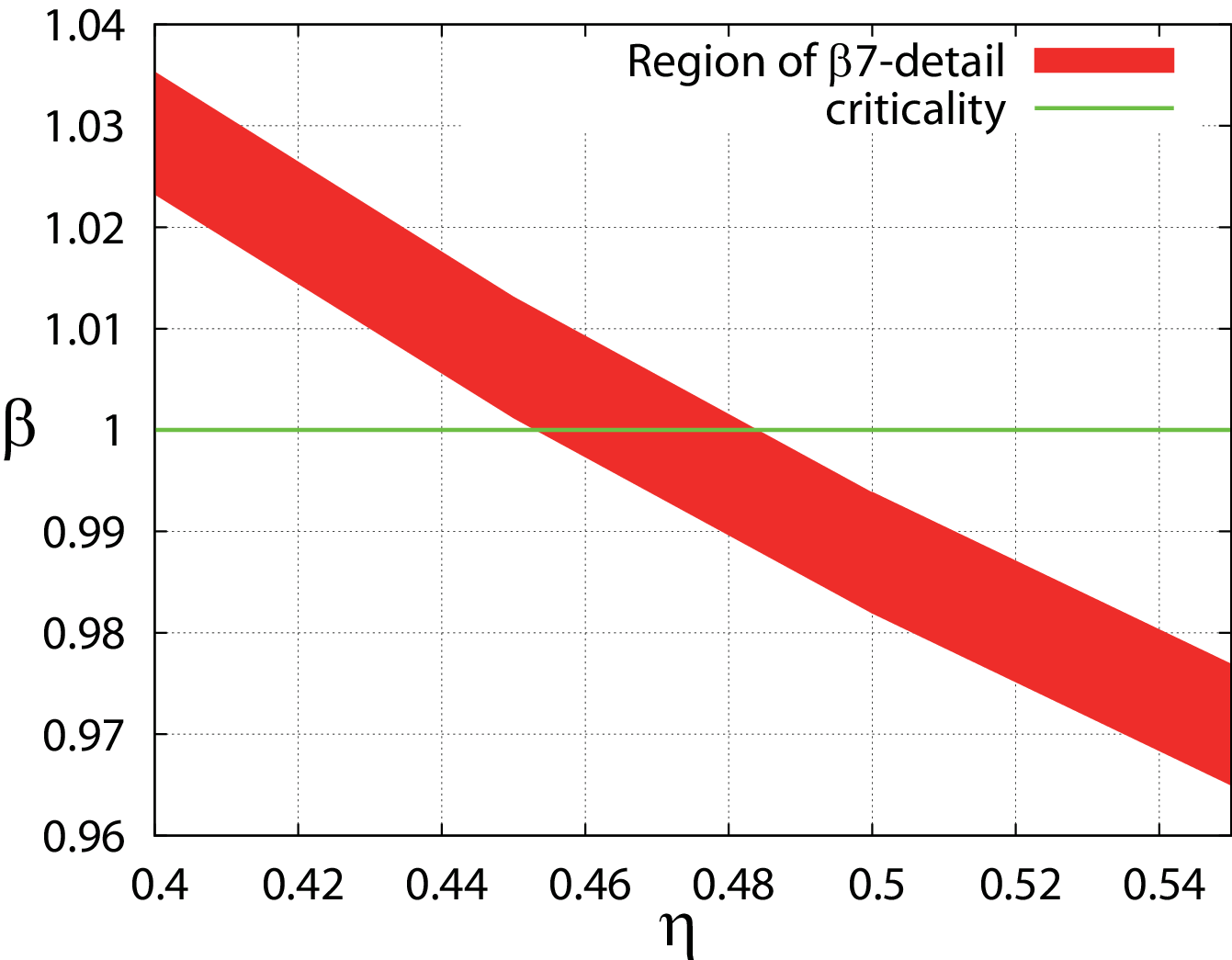}
\caption{$1\sigma$ region of $\beta$.}
\label{fig:beta-region}
\end{center}
\end{minipage}
\end{figure}

After checking the critical region in the above `rough' data, we increase the 
Monte Carlo data points to be 10.24 million and take 64 different set of random
numbers, which we call `detailed' data. 
In fig.\ref{fig:beta-detail}, the results of `detailed' $\beta$ are plotted 
with $1\sigma$ errors, which appears consistent with the `rough' data whose
$1\sigma$ region are expressed by curved lines.

Using the $1\sigma$ error for each $\eta$ point, we get $1\sigma$
region of $\beta$ estimate as in fig.\ref{fig:beta-region}. 
Then the crossing period of $\beta=1$ line 
with this region gives our estimate of $\eta_{\rm c}$.

\begin{figure}[ht]
\begin{minipage}{0.45\hsize}
\begin{center}
\includegraphics[width=0.99\textwidth]{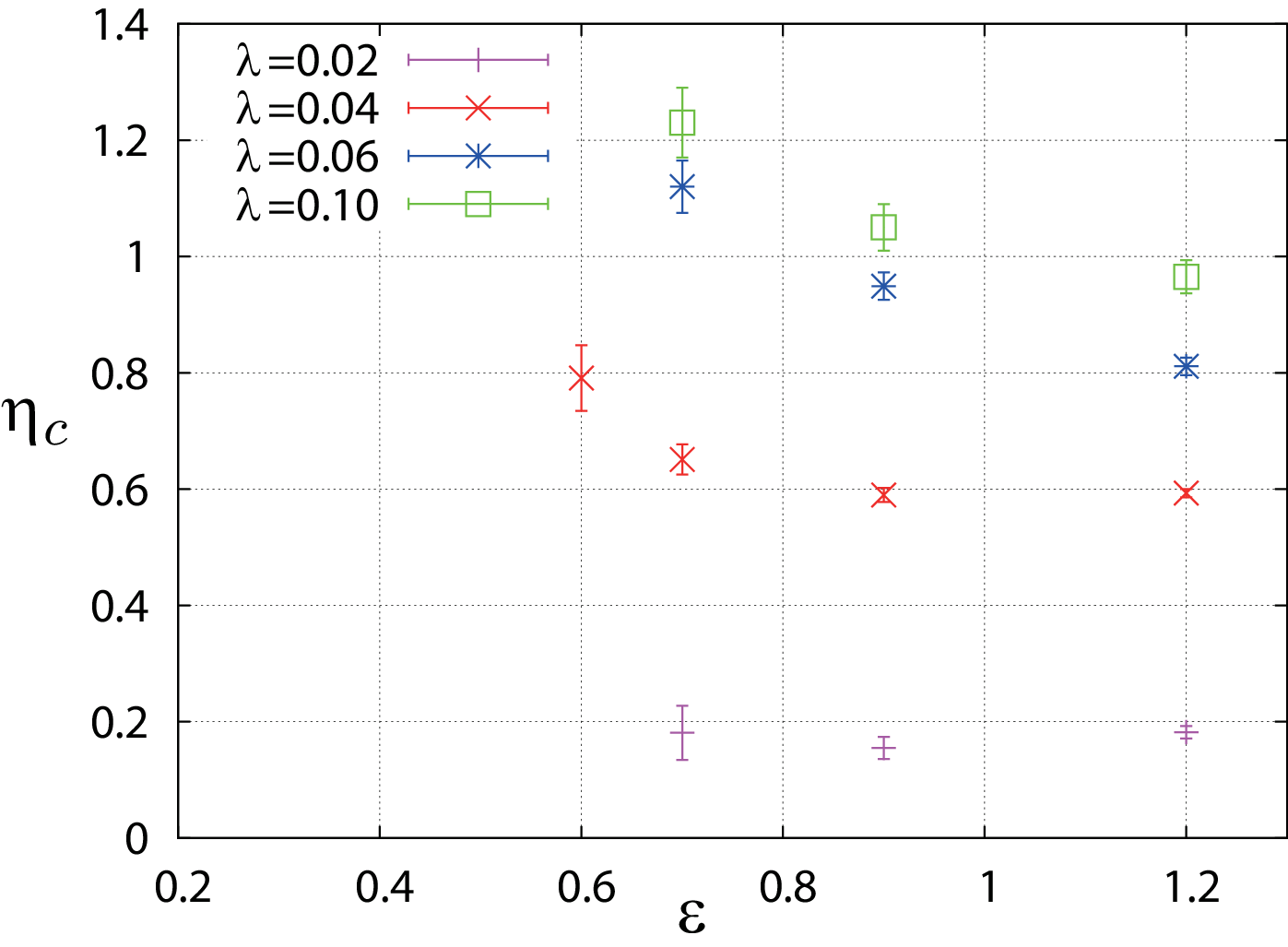}
\caption{$\epsilon$ dependence of $\beta$ ($n=6$)}
\label{fig:epsilon-dependence-n6}
\end{center}
\end{minipage}
\hfill
\begin{minipage}{0.45\hsize}
\begin{center}
\includegraphics[width=0.99\textwidth]{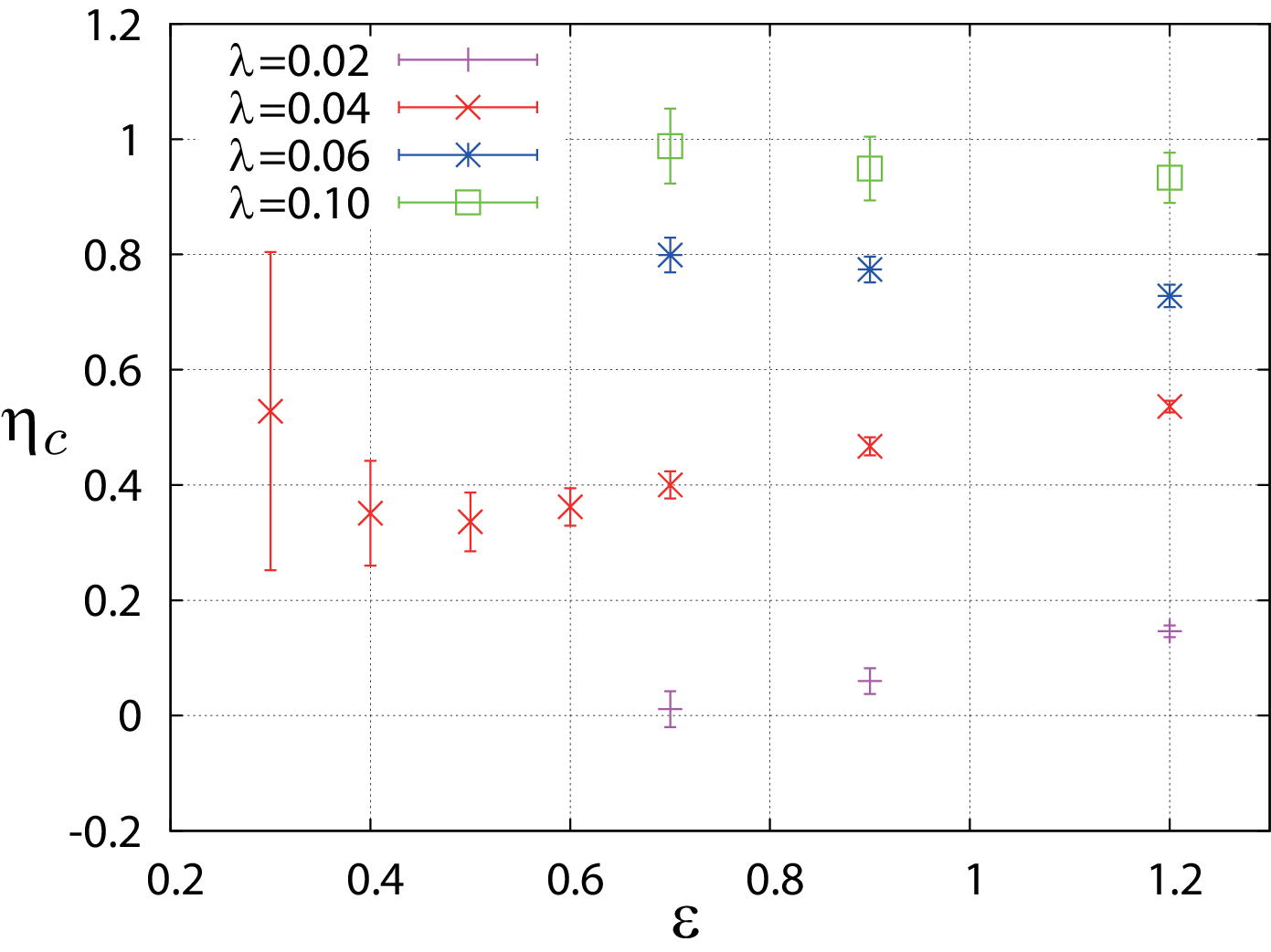}
\caption{$\epsilon$ dependence of $\beta$ ($n=7$)}
\label{fig:epsilon-dependence-n7}
\end{center}
\end{minipage}
\end{figure}

Next we check the $\epsilon$ dependence of $\eta_{\rm c}$.
The discretization skip $\epsilon$ should be small enough. 
However, the situation is not so simple and we cannot take 
the vanishing $\epsilon$ limit straightforwardly. 
The small $\epsilon$ means the length scale of the distance 
between spins becomes small, and the long range interaction nature of the 
system is being lost with fixed finite $n$. 
Actually we can calculate 
only rather short range interaction cases around $n=6,7$ because of
the computer resource limit.
Then the too-small $\epsilon$ finally breaks the FRS framework.

There must be an optimized finite $\epsilon$ which gives the best
value for the physical results within the limited calculational resources. 
Investigating the $\epsilon$ dependence of $\eta_{\rm c}$ and
size of their errors in fig.\ref{fig:epsilon-dependence-n6} ($p=1.99,\ n=6$) 
and fig.\ref{fig:epsilon-dependence-n7} ($p=1.99,\ n=7$),
we adopt $\epsilon=0.9 (n=6)$ and $\epsilon=0.7 (n=7)$ 
for major estimates with whole range of
parameter values, mainly taking account of total stability of calculations.

\begin{figure}[ht]
\begin{center}
\includegraphics[width=0.6\textwidth]{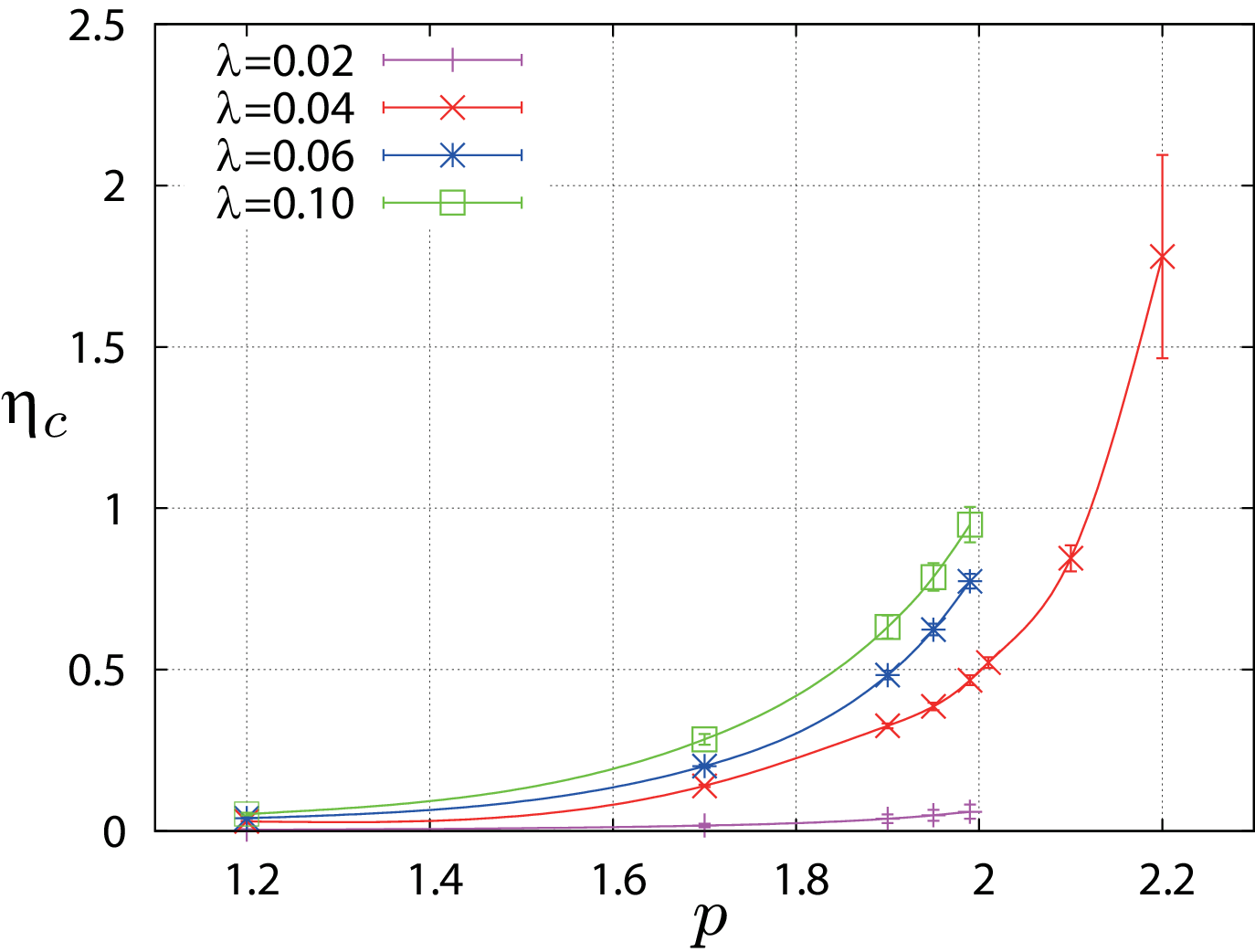}
\caption{$p$ dependence of $\beta$}
\label{fig:p-dependence}
\end{center}
\vskip-3mm
\end{figure}

Then we plot the $p$ dependence of $\eta_{\rm c}$ in 
fig.\ref{fig:p-dependence} ($\epsilon=0.9,\ n=7$).
Larger $p$ is equivalent to weaker interactions among 
spins. Therefore $\eta_{\rm c}$ should increase according
to $p$. The dependence resembles to the case of the normal long range Ising 
models\cite{Aoki-Kobayashi-Tomita08}.

However we observe still finite $\eta_{\rm c}$ in case of
large $p>2$, which is quite different from the normal long range 
Ising models defined in Eq.(\ref{eq:Ising}) where $\eta_c$ is proved to be infinite for $p>2$.
This issue of finite critical coupling constants seen for 
$p>2$ is an interesting problem indicating some new features
of the model or the important defects of our approximated model. 
Actually our model here does not correspond to a usual simple 
long range Ising model. Actually
we may define and evaluate 
effective Ising spin interactions given by the ground state
approximation. They are quite different from the simple-minded
interactions which are actually realized in case of the
delta-function states. There appear strong multi spin
($\sigma^4, \sigma^6$ etc.) interactions and enhanced long range
interactions.

In this sense it is non-trivial that
our approximation of the quantum double-well system may show up
the FRS behaviors just as in the normal long range Ising model,
and even more the quantum double-well
system with long range interactions has the same boundary structure
at $p=2$ as the normal long range Ising model.
To conclude something more definite about this issue we need to 
proceed beyond the two-state approximation 
where FRS will be examined with the full system.

\begin{figure}[ht]
\begin{center}
\includegraphics[width=0.75\textwidth]{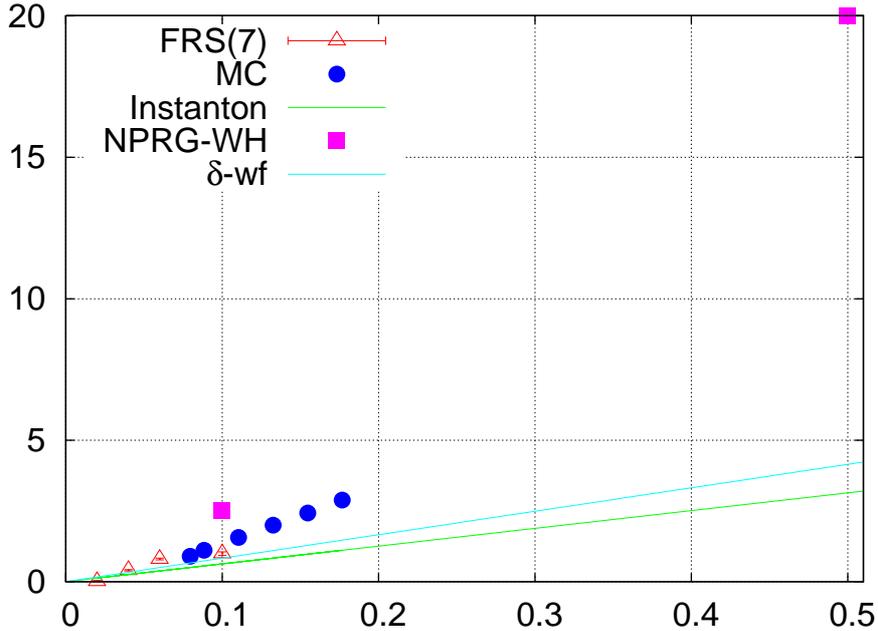}
\caption{Critical dissipation compared with other results.}
\label{fig:results}
\end{center}
\end{figure}

\begin{figure}[ht]
\begin{center}
\includegraphics[width=0.75\textwidth]{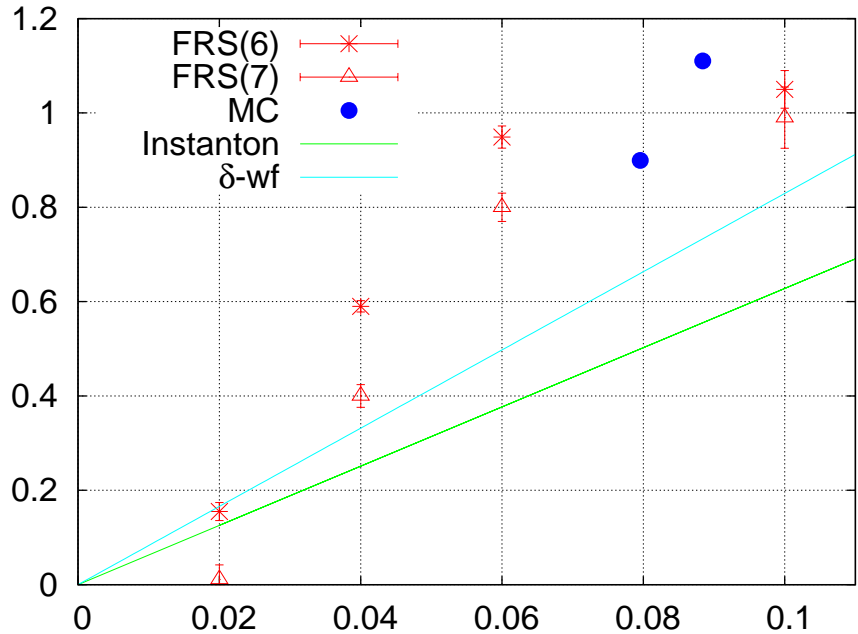}
\caption{Critical dissipation of small $\lambda$ region.}
\label{fig:results-zoom}
\end{center}
\end{figure}

To compare our results with earlier researches, we plot
$\eta_c$ versus $\lambda$ in Figs.\ref{fig:results} and \ref{fig:results-zoom}.
Here we take $p=1.99$, which is for the future comparison with normal long
range Ising calculation with FRS since $p=2$ there is hard to evaluate 
due to the boundary Kosterlitz-Thouless transition.
As for other parameters, we adopt two sets: $\epsilon=0.9, n=6$ 
(FRS(6) in figures) and $\epsilon=0.7, n=7$ (FRS(7) in figures).
We show data with error bars which show statistical errors because of the
Monte Carlo integration part in our method. 
We see some variation of results between these two sets of parameters, 
which indicate size of systematic errors
of our method of calculation in this stage.

Our FRS results and those of sophisticated Monte Carlo simulation 
(MC in figures)\cite{MC-DW} support complementary parameter regions.
At the larger $\lambda (>0.05)$ area, our results by 2-state approximation 
seems to deviate from the MC, 
and we consider that it reveals the insufficientness of our approximation of 
picking up only 2-state at each site in this region. The reason is that
at larger $\lambda$ region, the well becomes shallower and the energy gaps
between the 1st and the 2nd excited states are not so large as that between the ground
and the 1st excited states.

On the other hand, the dilute gas approximated instanton results 
(Instanton in figures)\cite{Instanton-DW}
look very much consistent with our results at small $\lambda (<0.02)$
region where both 2-state approximation and the dilute gas instanton 
are expected to work well there.
The dilute gas instanton calculation keeps only the position of instantons
and anti-instantons for configurations and neglect energy between 
instantons, that is, non-interacting gas. Then the dissipation effects are
evaluated for instanton configurations. This approach can be regarded as an 
another type of two-state approximation, and therefore its predictable
region resembles to ours.
Thus, the dilute gas instanton at larger $\lambda (>0.04)$ is not a good
approximation.

We plot the 2-state approximation results with the delta function 
wave function ($\delta$-wf in figures) 
explained in Eq.(\ref{eq:delta-wf}) though it suffers a 
serious problem for vanishing $\epsilon$ limit.
It is just above the instanton results and seems better than that.

Accordingly our method is a good candidate to give the critical coupling constants
in the intermediate region 
between the MC plausible region and the dilute gas instanton region.
However, it seems that within 2-state approximation, our method is not good enough to 
smoothly connect the MC and instanton. 
We have to proceed to multi-state or full-state
calculations, to which we will apply the Finite Range Scaling.

We also plot a non-perturbative renormalization group 
approach (NPRG-WH in figures)\cite{Aoki-Horikoshi} with the local potential
approximation, which 
gives results at very large $\lambda (>0.1)$ region. 
This approach uses the Wegner-Houghton equation where the effective 
Wilsonian action consists of the fixed kinetic term and the general 
potential term ignoring all other derivative interactions.
It is known to give very good results of the susceptibility
for large $\lambda (>0.1)$ case\cite{nprg-qm}. Ignoring the derivative terms in the
Wilsonian effective potential means it is not a good approximation 
for configurations with a sharp change of $x(\tau)$, and it fails
for small $\lambda (>0.1)$ region.
In total, it looks consistent with other approaches, though 
there is no other results in the large $\lambda (>0.1)$ region ever.

In the appendix we list our results of $\eta_{\rm c}$ for various
parameter values of $\lambda, p, \epsilon, n$. 

\section*{Acknowledgments}

@We thank fruitful discussions with Daisuke Sato and Kazuhiro Miyashita, and
also collaboration of Masakazu Arimoto, Yasuhiro Fujii and Hiroshi Tomita at the early stage of this work.

This research was partially supported by the Ministry of Education, Culture, Sports, Science and Technology 
through a Grant-in-Aid for Challenging Exploratory Research (No.12011251, 2012).

\section*{Appendix}

@We list the critical dissipation for various model parameters.
The $1\sigma$ errors are put in the parentheses.

\begin{center}
\centerline{Critical dissipation $\eta_{\rm c}$}
\vskip5mm
\begin{tabular}{|l|l|l|l|l|}
\hline
&\multicolumn{4}{c|}{$\lambda\ \  (\epsilon=0.9, n=6)$}\\
\cline{2-5}
p&0.02&0.04&0.06&0.10\\
\hline
1.2&0.01(0.00097)&0.041(0.0005)&0.057(0.0008)&0.069(0.0026)\\
1.7&0.044(0.0049)&0.18(0.002)&0.26(0.004)&0.32(0.012)\\
1.9&0.097(0.011)&0.37(0.007)&0.59(0.011)&0.70(0.024)\\
1.95&0.13(0.016)&0.48(0.009)&0.76(0.017)&0.87(0.031)\\
1.99&0.16(0.019)&0.59(0.012)&0.95(0.024)&1.05(0.040)\\
\hline\hline
&\multicolumn{4}{c|}{$\lambda\ \  (\epsilon=0.9, n=7)$}\\
\cline{2-5}
p&0.02&0.04&0.06&0.10\\
\hline
1.2&0.0039(0.0012)&0.030(0.0006)&0.040(0.0009)&0.053(0.0037)\\
1.7&0.017(0.0056)&0.14(0.0035)&0.20(0.005)&0.28(0.017)\\
1.9&0.038(0.013)&0.33(0.0075)&0.48(0.014)&0.63(0.036)\\
1.95&0.049(0.017)&0.39(0.012)&0.62(0.018)&0.79(0.043)\\
1.99&0.060(0.022)&0.47(0.016)&0.77(0.023)&0.95(0.055)\\
\hline
\hline
&\multicolumn{4}{c|}{$\lambda\ \  (\epsilon=0.7, n=6)$}\\
\cline{2-5}
p&0.02&0.04&0.06&0.10\\
\hline
1.99&0.18(0.047)&0.65(0.026)&1.12(0.045)&1.23(0.060)\\
\hline
\hline
&\multicolumn{4}{c|}{$\lambda\ \  (\epsilon=0.7, n=7)$}\\
\cline{2-5}
p&0.02&0.04&0.06&0.10\\
\hline
1.99&0.011(0.031)&0.4(0.024)&0.80(0.030)&0.99(0.065)\\
\hline
\end{tabular}
\end{center}

\end{document}